\documentclass[aps,prl,twocolumn,showpacs,showkeys]{revtex4}

\usepackage{graphicx,color,subfig,amsmath,bm,amssymb}

\begin{document}

\title{Strong electron-phonon coupling of the Fe breathing mode of
  LaO$_{1-x}$F$_x$FeAs} 

\author{Helmut Eschrig}

\email{h.eschrig@ifw-dresden.de}
\homepage{http://www.ifw-dresden.de/institutes/itf/members/helmut}

\affiliation{IFW Dresden, PO Box 270116, D-0111171 Dresden, Germany}

\begin{abstract}
  The electron-phonon coupling of LaO$_{1-x}$F$_x$FeAs is
  re-investigated on the basis of density functional theory in local
  density approximation. The implications of the $(\pi,\pi)$ nesting of
  the Fermi surfaces are carefully studied and found to lead to a
  non-standard electron-phonon coupling of the corresponding Fe in-plane
  breathing mode, which might make a strong contribution to the high
  superconducting transition temperature. The semi-metallic behavior of
  the undoped material is also further illuminated.
\end{abstract}


\maketitle

\begin{figure}
  \centering
  \includegraphics[height=0.9\columnwidth,angle=-90,clip=]{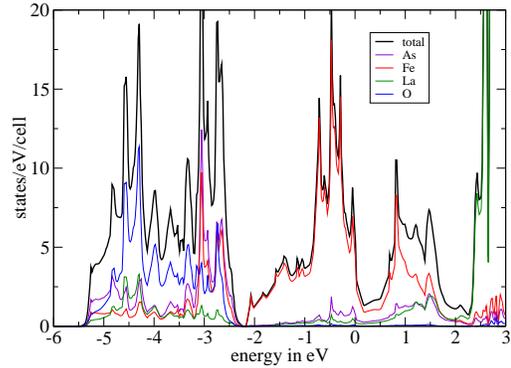}
  \caption{(color online) Total and component decomposed densities of
    states of LaOFeAs.} 
  \label{fig:f1}
\end{figure}

\begin{figure}
  \centering
  \includegraphics[height=\columnwidth,angle=-90,clip=]{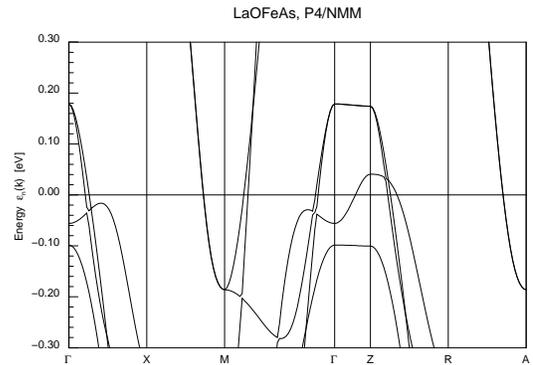}
  \caption{Bands close to the Fermi level ($\varepsilon_{\text{F}} = 0$)
    in the P4/nmm zone.}
  \label{fig:f5}
\end{figure}

The recently discovered \cite{kamihara08.1} layered superconductor
LaO$_{1-x}$F$_x$FeAs, $0.03 < x < 0.13$, attracts at present
considerable interest due to its comparably high critical temperature
$T_c \approx 40$K and possibly high upper critical field $H_{c2} \approx
50$Tesla, due to its similarities to the cuprates, and due to many
unusual properties. One peculiarity of all density functional based
electronic structure calculations
\cite{singh08.1,xu08.1,boeri08.1,mazin08.1,cao08.1,ma08.1} is that,
although there is strong Fe-$3d$ As-$4p$ covalency resulting in a total
width of the corresponding band complex of about 7eV, in a window of
about 2.5 eV around and below the Fermi level the electronic states are
of nearly pure Fe-$3d$ character (Fig.~\ref{fig:f1}). Mazin \textit{et
  al.}\ \cite{mazin08.1} pointed out that there is a strong Fermi
surface (FS) nesting in the undoped material with $q=(\pi,\pi,0)$
nesting wave vector between the two hole cylinders around the tetragonal
axis $\Gamma-Z$ and the two electron cylinders around the axis $M-A$
shifted by $q$ against $\Gamma-Z$ (Fig.~\ref{fig:f5}). This nesting led
those and several other authors
\cite{cao08.1,ma08.1,kuroki08.1,dong08.1} speculate on an important role
of antiferromagnetic correlations with magnetic in-plane superstructure
vector $(\pi,\pi)$, an apparently well justified scenario.
Electron-phonon (e-p) coupling obtained on the basis of density
functional perturbation theory \cite{singh08.1,boeri08.1} yielded only a
weak coupling quite not enough to explain the high $T_c$.

\begin{figure}
  \centering
  \includegraphics[width=0.6\columnwidth,clip=]{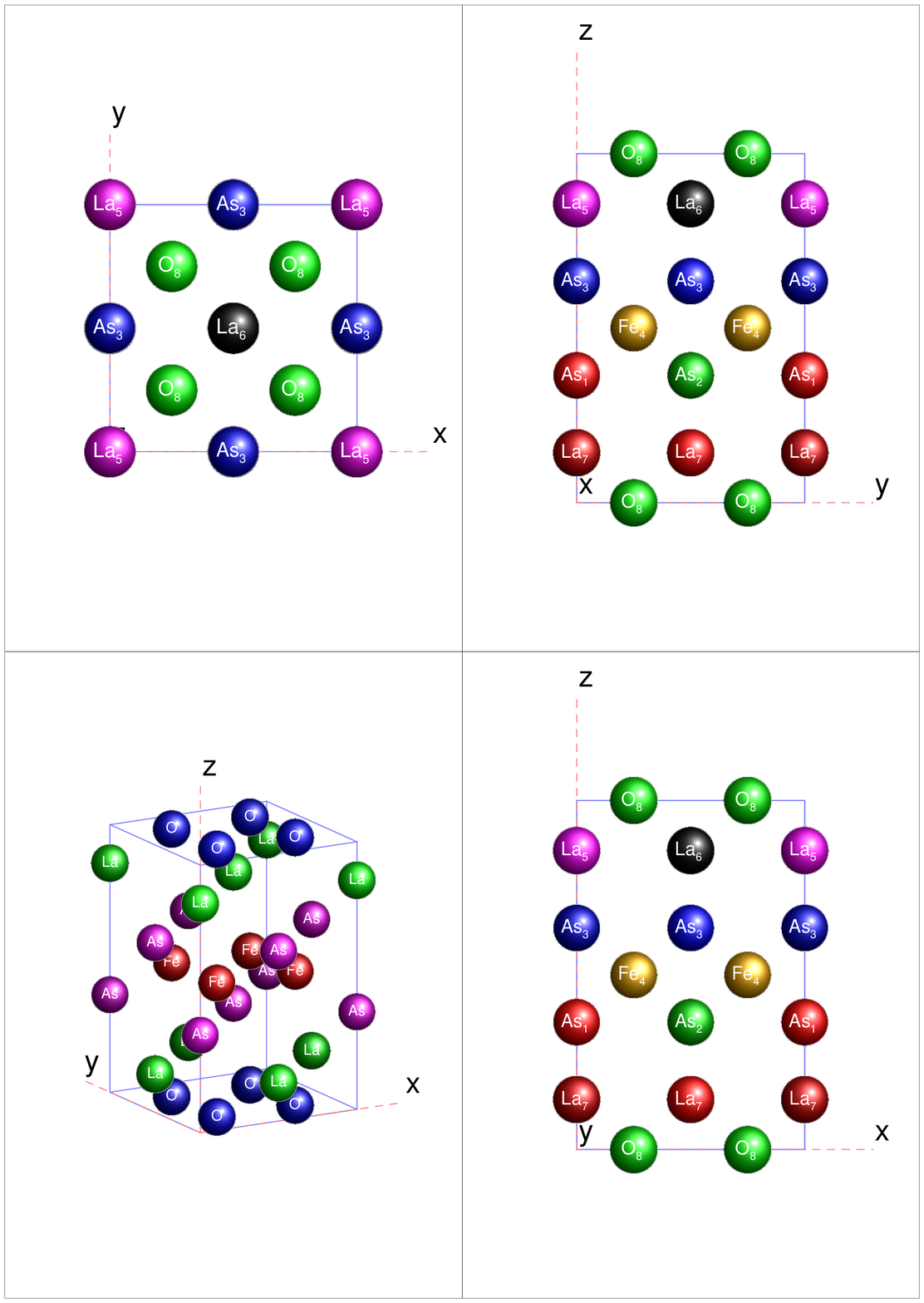}
  \caption{(color online) The P4mm unit cell (Fe breathing mode) of
    LaOFeAs.}
  \label{fig:f2}
\end{figure}

However, as is well known \cite{kohn59,migdal58} FS nesting also leads
to phonon anomalies, and the corresponding e-p coupling may not be
correctly described in lowest order perturbation theory which disregards
the degeneracy. Moreover, since metallicity and superconductivity take
place in the Fe layers of this quasi-two-dimensional tetragonal material
only (Fig.~\ref{fig:f2}), an average of linearized e-p coupling over the
whole structure might be misleading.  In the present text we report
results of investigation of these points which might be equally
important compared to the magnetic correlations to understand the
material. It is indeed found that e-p coupling may be quite strong. 

High precision density functional calculations were performed with the
local density approximation functional of \cite{perdew.92} using the
full-potential local-orbital code \cite{e134} in the version FPLO7-28
\cite{FPLO} with its default orbital settings. The experimental lattice
constants of undoped LaOFeAs, $a=4.03552$\AA, $c=8.7393$\AA, were used,
and the two Wyckoff parameters were optimized to be $z_{As}=0.365,\;
z_{La}=0.1435$.

\begin{figure}
  \centering
  \mbox{
  \includegraphics[width=0.3\columnwidth,clip=]{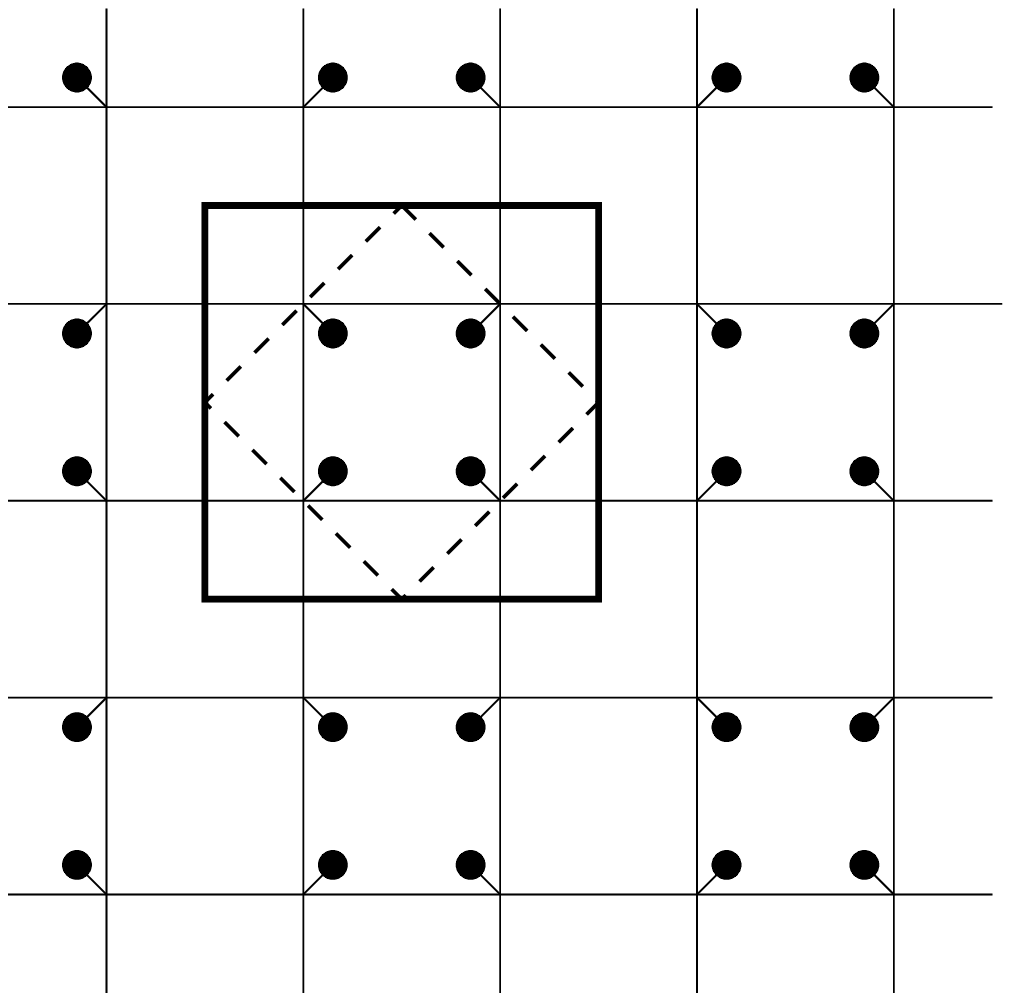}
  }
  \raisebox{0.35\columnwidth}{
  \includegraphics[height=0.55\columnwidth,angle=-90,clip=]{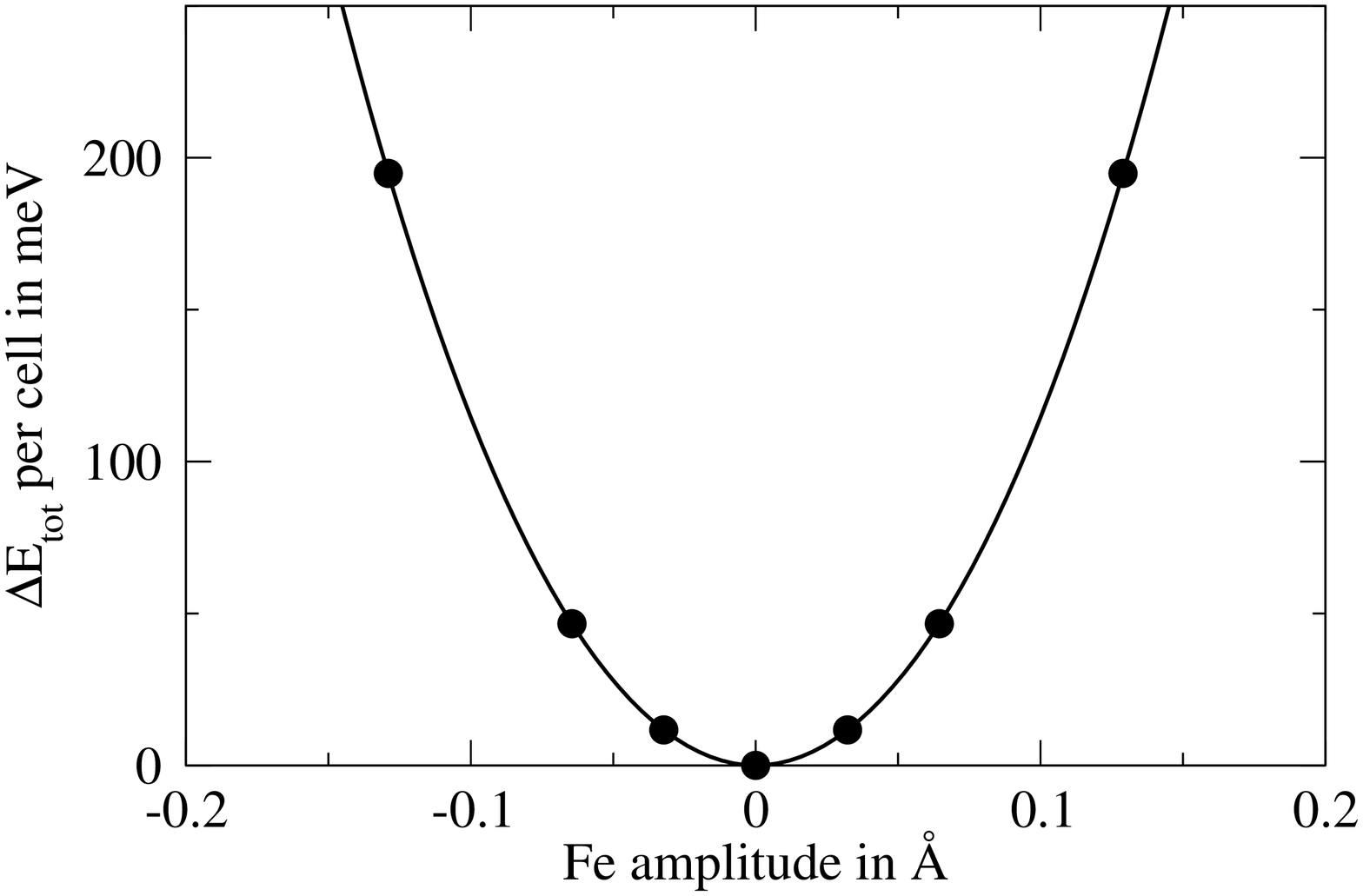}
  }
  \caption{Left:Displacement pattern of the Fe square lattice for the Fe
    breathing mode. The P4mm unit cell is shown (thick lines) and the
    P4/nmm unit cell (dashed lines, its center of inversion for the full
    structure is half-way between neighboring Fe sites).
    Right: Total energy per two formula units vs. Fe displacement of the
    Fe breathing mode.}
  \label{fig:f3}
\end{figure}

\begin{figure}
  \centering
  \includegraphics[height=\columnwidth,angle=-90,clip=]{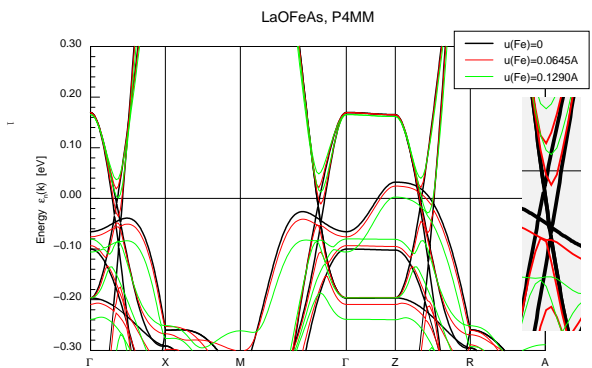}
  \caption{(color online) Band deformations due to the Fe breathing mode
    in the P4mm zone. The upward dispersing band on the line $\Gamma -
    M$ is two-fold degenerate. The inset on the right shows the zoomed
    detail from the line $M-\Gamma$.}
  \label{fig:f6}
\end{figure}

The phonon mode relevant in this respect is the in-plane breathing mode
of the Fe square lattice (checkerboard pattern where every other white
square is shrunk, Fig.~\ref{fig:f3}, left part). A static displacement
of Fe according to this breathing mode reduces the P4/nmm symmetry of
the lattice to P4mm by doubling the unit cell and removing the inversion
from the space group P4/nmm. (The center of inversion was half-way
between two neighboring Fe sites.) Doubling the unit cell of the lattice
means folding down the unit cell of the reciprocal lattice, the
Brillouin zone. The original point $M: \bm q = (\pi,\pi,0)$ of the
P4/nmm lattice (Fig.~\ref{fig:f5}) is folded onto $\Gamma: \bm q' =
(0,0,0)$ of the P4mm lattice; the folded band structure from
Fig.~\ref{fig:f5} is shown on Fig.~\ref{fig:f6}. For symmetry reasons,
at $\bm q=(\pi,\pi,q_3)$ the Fe-mode shown on Fig.~\ref{fig:f3} could
only mix with a corresponding oxygen breathing mode; both are, however,
expected to interact only very weakly, and we neglect this mixing. The
total energy change due to a static displacement (frozen phonon) is
shown on Fig.~\ref{fig:f3}, right part.  Despite the FS nesting the mode
potential is quite harmonic, the reason for this quadratic dependence on
the Fe displacement $u_{\text{Fe}}$ is given below. It corresponds to a
frequency of $\omega_{\text{Fe}} = 29.4$meV in good agreement with the
upper limit of the metal mode energies obtained in \cite{singh08.1}.
Note that there are two equivalent such modes of the Fe atoms as a
rotation of the displacement vector by $90^\circ$ in the left part of
Fig.~\ref{fig:f3} corresponds just to an in-plane shift of the whole
P4mm lattice by $(1/2,0,0)$ which together with a reflection of the
$c$-axis is a glide symmetry element of the original P4/nmm lattice.

\begin{figure}
  \centering
  \includegraphics[height=\columnwidth,angle=-90,clip=]{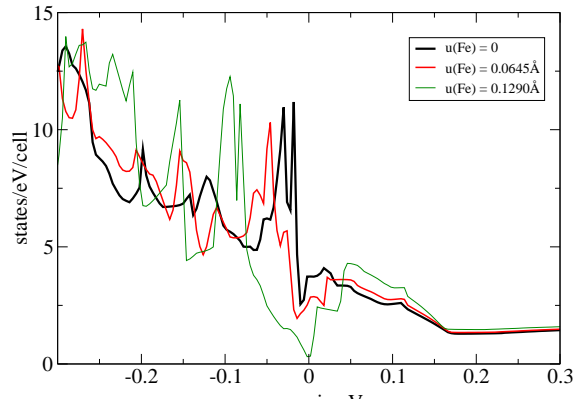}
  \caption{(color online) Deformation of the total DOS due to the Fe
    breathing mode (states per two formula units).} 
  \label{fig:f7}
\end{figure}

\begin{figure}
  \centering
  \includegraphics[height=\columnwidth,angle=-90,clip=]{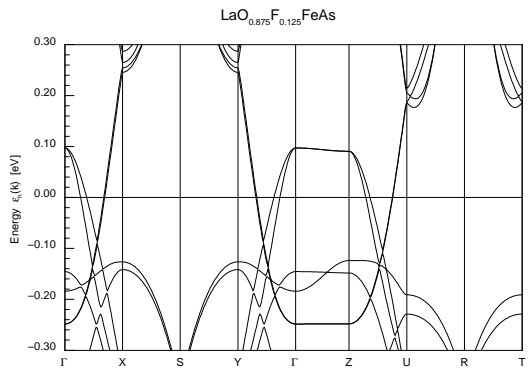}
  \caption{Bands of LaO$_{0.875}$F$_{0.125}$FeAs in the Pmm2 zone.}
  \label{fig:f8}
\end{figure}

On Fig.~\ref{fig:f6} the bands corresponding to the undistorted lattice
and to two static mode amplitudes, respectively, are shown in the
downfolded P4mm zone. The common Fermi level is again
$\varepsilon_{\text{F}}=0$. Now, the nesting for $\bm q'=0$ is seen as
band crossings very close to the Fermi level (observe the energy scale
of the figure) on the lines $\Gamma-M$ (and $\Gamma-X$) and $Z-R$ (the
latter for $q'_3=\pi$; the new point $M$ of the folded zone corresponds
to point $X$ on Fig.~\ref{fig:f5}.) From these bands as function of the
displacement $u_{\text{Fe}}$ of the Fe atoms, deformation potentials are
derived.  Care is, however, needed since deformation potentials are
touchy entities \cite{khan.84}. First of all, they have to be determined
from a common Fermi level. In the present case, moreover, reversion of
$u_{\text{Fe}}$ is equivalent to an in-plane shift of the whole P4mm
lattice by $(1,1,0)$ which is a lattice vector of the original P4/nmm
lattice, and hence changes in band energies $\Delta\varepsilon_{\bm k
  n}$ must be even in $u_{\text{Fe}}$.  Consequently, well away from
band crossings there is no first order deformation potential,
$\Delta\varepsilon_{\bm k n} \sim u^2_{\text{Fe}}$. At crossing of
folded bands according to the symmetry of the lattice with
$u_{\text{Fe}} \neq 0$ present (Fig.~\ref{fig:f6}), gaps open with
$\Delta\varepsilon_{\text{gap}} \sim |u_{\text{Fe}}|$. In the vicinity
of those crossings one has
\begin{equation}
  \label{eq:1}
  \Delta\varepsilon_{\bm k n} \approx 
  \pm\sqrt{(v\Delta\bm k)^2 + (I_{\text{Fe}} u_{\text{Fe}})^2}
  \mp |v\Delta\bm k|,
\end{equation}
where $v$ is the average of the two band velocities at the crossing,
$\Delta\bm k$ is the deviation from the crossing point, and
$I_{\text{Fe}} u_{\text{Fe}}$ is the deformation potential at the
gap:
\begin{equation}
  \label{eq:2}
  \Delta\varepsilon_{\text{gap}} \approx 2I_{\text{Fe}}u_{\text{Fe}}.
\end{equation}
For the displacement corresponding to one phonon per unit cell with
energy $\omega_{\text{Fe}}, \; \langle u^2_{\text{Fe}}\rangle =
(2M_{\text{Fe}}\omega_{\text{Fe}})^{-1/2}$ one has $I_{\text{Fe}}\langle
u^2_{\text{Fe}}\rangle^{1/2} \approx 20$meV$ \approx
0.7\omega_{\text{Fe}}$. By expanding the square root of (\ref{eq:1}) for
$|v\Delta\bm k| = 20$meV, approximately the position of the FSs of the
fluor doped material (at lower doping level compared to
Fig.~\ref{fig:f8}, where 12.5 p.c. fluor were considered to limit the
computational effort), an FS-averaged e-p coupling
constant
\begin{equation}
  \label{eq:3}
  g = \overline{g_{\bm kn,\bm k'n'}} = \frac{1}{2}\,
  \frac{I^2_{\text{Fe}} \langle u^2_{\text{Fe}}\rangle}{20\text{meV}}
  \approx 10\text{meV} \approx 
  I_{2,\text{Fe}}\langle u^2_{\text{Fe}}\rangle
\end{equation}
is obtained, where $I_{2,\text{Fe}}$ is the corresponding second order
deformation potential. Of course, whenever a phonon mode connects two
points on the FS, this gaping with lattice displacement happens.
However, normally it happens just for those two points without
statistical weight. The peculiarity here is that, as a consequence of the
nesting, for one mode it happens in the vicinity of the whole FS in a
layer of thickness $\sim\omega_\nu$ in energy (cf. Fig.~\ref{fig:f7}).

In the standard theory the e-p coupling strength $\lambda = Z(0) - 1 =
i(\partial\Sigma/\partial\varepsilon)_{\varepsilon = 0}$ is obtained
from the e-p self-energy
\begin{equation}
  \label{eq:4}
  \Sigma(\varepsilon) = \sum_{n\nu}\int\frac{d\omega}{2\pi}
  \int\frac{d^3k}{(2\pi)^3} g^2 
  \frac{2\omega_\nu}{\omega^2 + \omega^2_\nu} 
  G_n(\varepsilon-\omega,\bm k)
\end{equation}
($n$ being the band index and $\nu$ the phonon mode index) by treating
the phonon quantities as nearly $\bm k$-independent and using the $\bm
k$-integrated electron Green's function
\begin{multline}
  \label{eq:5}
  \sum_n \int\frac{d^3k}{(2\pi)^3} G_n(\varepsilon - \omega,\bm k) =\\
  = N(0) \int_{-\varepsilon_c}^{\varepsilon_c} d\varepsilon_c 
  \frac{1}{i(\varepsilon - \omega) - \varepsilon_c} =\\
  = -i\pi N(0)\text{sign}(\varepsilon - \omega) + O(1/\varepsilon_c),
\end{multline}
where $N(0)$ is the electron density of states (DOS) at the Fermi level,
and the high-energy cutoff $\varepsilon_c$ is considered negligible
small (of the order of Migdal's parameter). Then, the derivative of sign
yields a delta function and the $\omega$-integration is readily done
yielding $\lambda = \sum_\nu 2N(0)g^2/\omega_\nu$.

In the present case, which is not subject to Migdal's theorem, this
approach is not allowed since the renormalized electron Green's
functions have a low energy structure at $\eta_c \approx 20$meV, the
distance of the gap edges from Fermi level, which causes additional
terms in (\ref{eq:5}). Moreover, Eq.~(\ref{eq:3}) represents a
two-phonon vertex (four-leg vertex), and the corresponding contribution
to the electron self-energy has two phonon propagators connecting the
two vertices and an additional $\omega$-integration in the self-energy,
and without the situation with the gaps one would again be left with
$\lambda \approx \sum_\nu N(0)g^2/\omega_\nu$, where only $g$ is now the
coupling strength (\ref{eq:3}) of the four-leg vertex, which in the
present case is of the same order of magnitude as the usual $g$ of an
ordinary three leg vertex. However, due to the gap structure all
frequency integrations are entangled, and resonances of the type
$1/(\omega_\nu - \eta_c)$ may appear in the course of self-consistency
for $G_n$ \cite{newns.07}. A quantitative theory for this case is
complicated and not yet well developed. Non of the approximate
assumptions leading to (\ref{eq:4}) and (\ref{eq:5}) can be maintained.
So far one could only speculate on the magnitude of the effects.

Now, the nesting is considered in more detail. A closer look on
Fig.~\ref{fig:f6} reveals that a displacement $u_{\text{Fe}} =
0.0645$\AA, which does not much exceed the average displacement of one
phonon per mode, already removes two of the four FSs of the in-plane
bands (Fe-$3d_{xz}$ and Fe-$3d_{yz}$ orbital character) by opening a gap
of about 75meV. Only an electron FS of the shape of a cylindric shell
around the tetragonal axis, quite isotropic perpendicular to it,
survives (and the hole pocked around point $Z$ of the band of
Fe-$3d_{z^2}$ orbital character). Just for illustration the effect of a
(not quite realistic) two times larger displacement is also plotted on
Fig.~\ref{fig:f6} for which the material becomes a semi-metal with only
two tiny FS pockets, an electron ring and a hole pocket, both around
$Z$. Even slightly larger displacements would make it a true
semiconductor. This is reflected in the behavior of the total DOS shown
on Fig~\ref{fig:f7} (where the normalization is for the unit cell of the
undistorted P4/nmm structure). One important point is that the nesting
is only approximate and, more importantly, that at a given energy the
bands which were originally electron or hole bands move in the same
direction. Since the band crossings are not exactly at Fermi level and
in e-p coupling theory the Fermi level has to be kept fixed
\cite{khan.84}, the considered e-p coupling is transmitted to all bands
and is also transmitted to the band of Fe-$3d_{z^2}$ character. These
findings (together with the similar magnetic peculiarities
\cite{mazin08.1}) are capable of explaining many of the unusual
electronic properties of the undoped material.

\begin{figure}
  \centering
  \includegraphics[height=0.8\columnwidth,angle=-90,clip=]{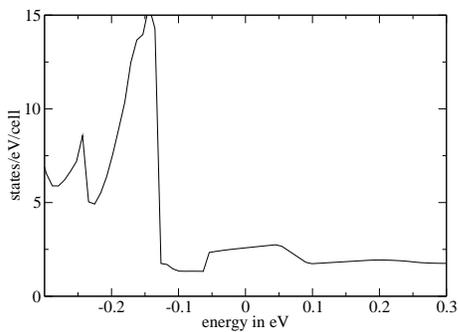}
  \caption{Total DOS of ordered LaO$_{0.875}$F$_{0.125}$FeAs (states per
    two formula units).}
  \label{fig:f9}
\end{figure}

Finally, the effect of fluor doping for oxygen is studied by replacing
one of eight oxygen atoms with a fluor atom in an ordered way. This
reduces the symmetry of the lattice further to the orthorhombic Pmm2
group with a four times larger unit cell, square in plane, compared to
the P4/nmm structure. The corresponding bands (again folded, into the
Pmm2 zone, which however does not results in new shifts of the FSs which
are already close enough to the central line $\Gamma - Z$) and total DOS
are shown on Figs.~\ref{fig:f8} and \ref{fig:f9}. As can be seen by
comparing Figs.~\ref{fig:f6} and \ref{fig:f7} with Figs.~\ref{fig:f8}
and \ref{fig:f9}, one fluor atom donates just one electron without
changing the Fe derived $t_{2g}$-bands $(d_{xz},\;d_{yz})$ in the
vicinity of the Fermi level.  Therefore, also no disorder effect of
fluor doping is expected in the relevant electronic structure. Moreover,
the Fe-$3d_{z^2}$ derived bands, which shift a little bit due to the
change of layer charges, are now completely filled (one of them
producing the DOS peak at about -150meV), and the corresponding hole
pocket around $Z$ has gone, leaving only the four slightly deformed
cylindrical FSs, two larger electron cylinders around the $\Gamma-Z$
axis and two smaller hole cylinders around the $M-A$ axis of the
original P4/NMM structure (cf.\ Fig.~\ref{fig:f5}).

In summary, we have shown that the e-p coupling in LaO$_{1-x}$F$_x$FeAs
is indeed very strong due to the Fe in-plane breathing mode, and this
cannot be treated in standard e-p theory of superconductivity. With
respect to this breathing mode, the material rather compares with
MgB$_2$, although the details are quite different, and the magnetic
couplings already discussed in the literature must also be taken
seriously into account. In that respect, the material may have much in
common with the cuprates. A principally similar but in detail quite
different case is a breathing mode of the copper square lattice in
connection with the approximate $(\pi,\pi)$-nesting \cite{newns.07}. It
is intriguing hence, whether LaOFeAs could provide some bridge between
the other two cases, MgB$_2$ and the cuprates. The undoped material
appears to be a bad metal with a great complexity of very strong
couplings, e-p and magnetic.  Doping (potentially both electron or hole)
moves LaOFeAs away from this messy coupling situation into a still
strong coupled but good metallic regime.  It is to be expected from
former \cite{mazin08.1} and the present analysis that LaOFeAs will be
more symmetric with respect to both electron and hole doping than the
cuprates.

\begin{acknowledgments}
  Helpful discussion with Klaus Koepernik, Stefan-Ludwig Drechsler, B.
  B\"uchner and in particular with M. Eschrig are gratefully
  acknowledged. I.I. Mazin pointed out to me a substantial error in an
  earlier version of the paper.
\end{acknowledgments}

\bibliography{../../bib/sc.bib,../../bib/part.bib,../../bib/eschrig.bib,%
../../bib/dft.bib}

\end{document}